\documentclass[referee]{raa}
\usepackage{graphicx,times}
\usepackage{natbib}
\usepackage{amssymb,amsmath}
\usepackage{booktabs}
\usepackage[colorlinks=true,pdfstartview=FitH,linkcolor=blue,anchorcolor=violet,citecolor=magenta]{hyperref}

\usepackage{array}
\usepackage{orcidlink}
\usepackage[utf8]{inputenc}
\usepackage[T1]{fontenc}
\usepackage{CJKutf8}

\newcommand{\songti}[1]{\begin{CJK*}{UTF8}{gbsn}#1\end{CJK*}}
\newcommand{\kaiti}[1]{\begin{CJK*}{UTF8}{gkai}#1\end{CJK*}}

\begin{document}

\title{Determining the observational epoch of the Shi's star catalog using the generalized Hough transform method}

\author{Boliang He (\kaiti{何勃亮}~\orcidlink{0000-0002-3244-7312}) \inst{1,2}
    \and Yongheng Zhao (\kaiti{赵永恒}~\orcidlink{0000-0001-5298-2833}) \inst{1,2}
    }
    
\institute{
  National Astronomical Observatories, Chinese Academy of Sciences, Beijing, 100101, China; {\it hebl@nao.cas.cn}
  \and 
  University of Chinese Academy of Sciences, Beijing, 100049, China; 
}

\abstract{
    Ancient stellar observations are a valuable cultural heritage, profoundly influencing both cultural domains and modern astronomical research. The \textit{Shi's Star Catalog} (\kaiti{石氏星经}), the oldest extant star catalog in China, faces controversy regarding its observational epoch. Determining this epoch via precession assumes accurate ancient coordinates and correspondence with contemporary stars, posing significant challenges. This study introduces a novel method using the Generalized Hough Transform to ascertain the catalog's observational epoch. This approach statistically accommodates errors in ancient coordinates and discrepancies between ancient and modern stars, addressing limitations in prior methods. Our findings date the \textit{Shi's Star Catalog} to the 4th century BCE, with 2nd-century CE adjustments. In comparison, the Western tradition's oldest known catalog, the Ptolemaic Star Catalog (2nd century CE), likely derives from the Hipparchus Star Catalog (2nd century BCE). Thus, the \textit{Shi's Star Catalog} is identified as the world's oldest known star catalog. Beyond establishing its observation period, this study aims to consolidate and digitize these cultural artifacts. 
}


\authorrunning{B.-L. He, Y.-Z. Zhao}
\titlerunning{Determination of the observational epoch of the Shi's star catalog}

\maketitle

\section{Introduction}\label{sec:intro}

Throughout human history, the night sky has persistently captivated collective fascination, with stars functioning as both a wellspring of inspiration and a locus of intellectual exploration. Each culture has crafted its distinctive celestial landscape, and over millennia, individuals have meticulously documented their perceptions of the night sky. The constellations and the nomenclature of stars have transcended mere astronomical entities to become emblematic cultural symbols within diverse civilizations. Amidst these multifaceted observational endeavors, scientific inquiry has emerged, culminating in unique and richly detailed records of stellar observations within both Eastern and Western traditions.

In the Western tradition, Ptolemy's seminal astronomical work, \textit{Almagest}, dating to the 2nd century CE, meticulously documented a star catalog that has endured to the present day. The epoch of this catalog is posited to be 137 CE \citep{2012A&A...544A..31V}. The Ptolemaic star catalog stands as the oldest extant star catalog in the Western world. Conversely, in China, the \textit{Shi Shi Xing Jing} \kaiti{石氏星经} (commonly referred to as \textit{Shi's} Star Manual) was composed by the astronomer \textit{Shi Shen} (\kaiti{石申}) during the Zhanguo (Warring States) period, spanning the 5th to 3rd centuries BCE. This seminal work encompasses a star catalog known as the \textit{Shi's Star Catalog}. Subsequently, the \textit{Shi's Star Catalog} was referenced in the Tang Dynasty (7th to 10th century CE) compilation \textit{Kaiyuan Zhanjing} (Kaiyuan Astrological Texts, \kaiti{开元占经}), from which modern scholars have excerpted the \textit{Shi's Star Catalog}.

The \textit{Shi's Star Catalog} meticulously records observational data for the coordinates of 120 stars. Over the past century, astronomers have primarily utilized retrograde precession to ascertain its observational epoch. This methodology entails cross-referencing the ancient catalog with contemporary star catalogs, followed by rigorous analytical procedures, ultimately leading to the definitive determination of the \textit{Shi's} observational epoch. Currently, four predominant theories prevail regarding the observational epoch of the \textit{Shi's Star Catalog} \citep{hisastro2015}:

\subsection*{(1) The 4th Century BCE Period}

Based on the records of precession and retrograde precession, it is inferred that the observational epoch of the \textit{Shi's Star Catalog} approximately corresponds to the Zhanguo period, around 360 BCE \citep{xincheng1926}.

\subsection*{(2) Dual Epochs: 5th Century BCE and 2nd Century CE}

Two interpretations prevail regarding the observational epochs of the \textit{Shi's Star Catalog}. One posits that 44 stars were observed circa 360 BCE, while 46 stars were documented around 200 CE, as determined through graphical methods \citep{1930PKwaO...1...17U}. Conversely, another perspective suggests that 52 stars were recorded circa 440 BCE, with an additional 37 stars observed around 160 CE, utilizing the conventional retrograde precession method \citep{PanNai2009HCSO}.

\subsection*{(3) The 1st Century BCE Period}

This viewpoint garners substantial scholarly support, with many asserting that the observational timeframe likely spans from 100 BCE to 70 BCE. The methodologies predominantly employed in this analysis include modern statistical methods and Fourier analysis \citep{qianbaocong1937,Yabuuchi1984,yasukatsu1977oldest,guoshengchi1994,sunxiaochun1994ssxg,sun1997chinese}.

\subsection*{(4) The 7th Century CE Period}

Some scholars also propose that the observations recorded in the \textit{Shi's Star Catalog} might have taken place during the early Tang Dynasty. However, this analysis remains insufficiently comprehensive \citep{huweijia1998}.

Therefore, a multitude of perspectives exist regarding the observational timeframe of the \textit{Shi's Star Catalog}, with no definitive consensus reached. The author of this study has previously employed the generalized Hough transform method to scrutinize star observation records from the Song and Yuan dynasties in China \citep{hebl2024013}, effectively ascertaining the observation epochs of these star catalogs. This paper serves as a continuation of that endeavor, aiming to further analyze the data within the \textit{Shi's Star Catalog} to determine its genuine observational epoch.

\section{Method}

In ancient China, astronomers employed a traditional equatorial coordinate system, which consisted of two principal components: \textit{ru xiu du} (RXD, the angular distance to the western boundary of a lunar mansion) and \textit{qu ji du} (QJD, the angular distance from the North Celestial Pole). The Chinese constellation system segmented the celestial sphere into 28 lunar mansions, each demarcated by a principal star (ju xing) that served as a reference. RXD quantified the longitudinal separation between a star and its corresponding principal star, whereas QJD denoted the star's angular distance from the North Celestial Pole. In contrast, contemporary astronomy utilizes the equatorial coordinate system, characterized by Right Ascension ($\alpha$) and Declination ($\delta$).

$$\textbf{s} = (\alpha, \delta)$$

The relationship between \textsf{QJD} ($\omega$) and the Declination ($\delta$) in the modern equatorial coordinate system is given by:

$$\delta = 90^{\circ} - \omega$$

The utilization of the Generalized Hough Transform method to ascertain the observational epochs of stars constitutes an innovative computational technique. This method merges a probabilistic model with the Monte Carlo algorithm, facilitating the calculation of parameters by leveraging the probability density within the parameter space. As a result, it accounts for potential inaccuracies in the precession values of ancient star catalogs and the ambiguities in the alignment between ancient and contemporary stars, provided that the predominant portion of the data remains valid, thus yielding dependable outcomes. Moreover, this method uniquely relies solely on QJD data for its analysis, distinguishing itself from other methodologies that require the conjunction of both \textsf{QJD} and \textsf{RXD} data.

The original dataset consists of rigorously validated stellar parameters, including essential attributes such as right ascension, declination, proper motion, and \textsf{QJD}, all referenced to the J2000.0 celestial coordinate system. These parameters are obtained from \textit{The Chinese Ancient Astronomy Fundamental Reference Star Catalog} \citep{xinghan.100877}. The catalog's data is sourced from the relatively precise star data of the Hipparchus Catalog \citep{1997ESASP1200.....E,2007A&A...474..653V,2012AstL...38..331A}, supplemented with recent updates and additional stellar parameters such as magnitude, proper motion, parallax, and radial velocity.

In the computational process, the positional parameters of the North Celestial Pole (right ascension and declination) are utilized as the primary focus for resolution within the parameter space. As a result, each star's representation within this space is depicted as a circle with its coordinates at the center and \textsf{QJD} as the radius. The selection of any two stars may yield intersections within the parameter space, which can be classified into three scenarios: no intersection, a single intersection, or two intersections.
 
\begin{figure}[htbp]
    \centering
    \includegraphics[width=0.5\textwidth]{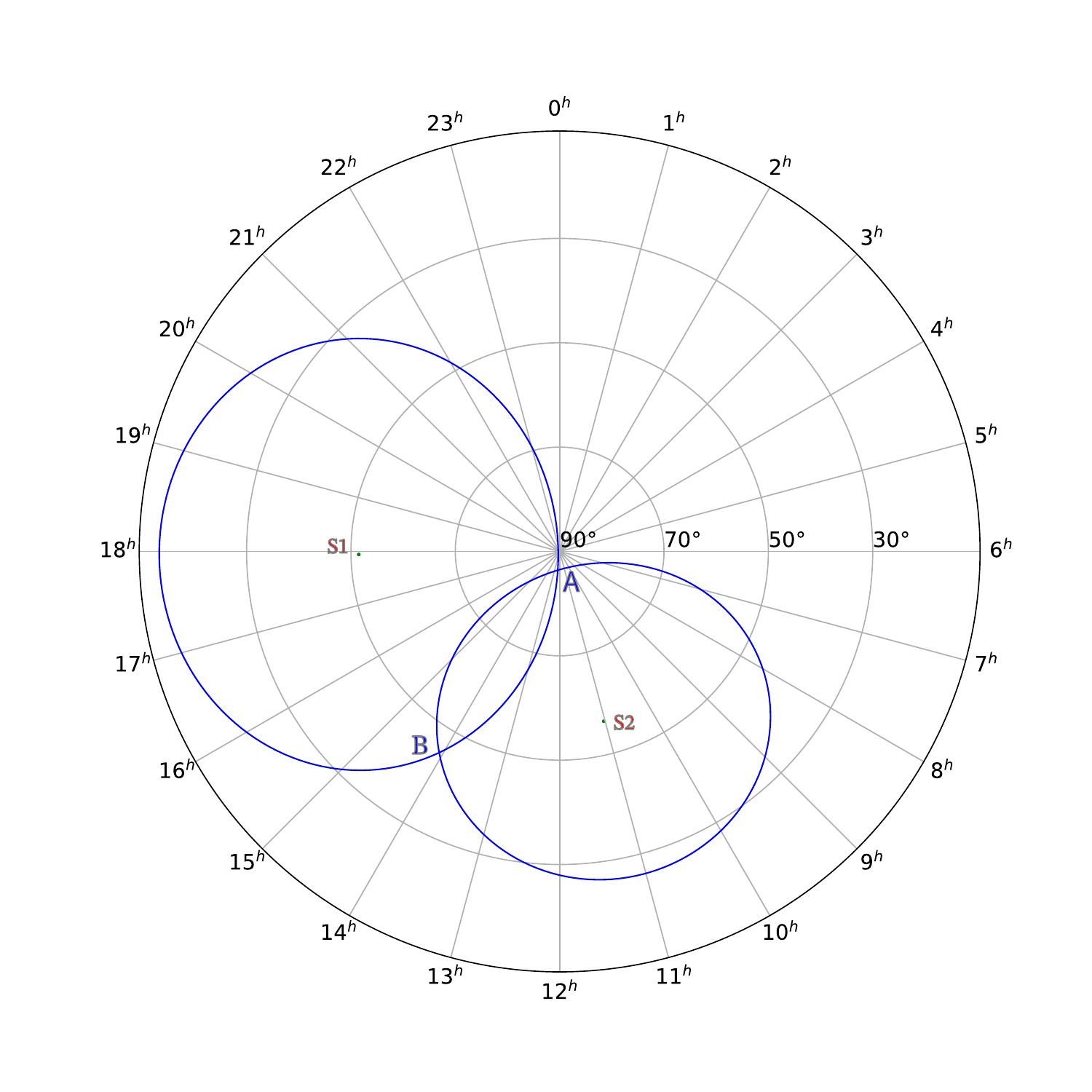}
    \caption{Simulation of the parameter space of the North Celestial Pole after generalized Hough transform.}\label{fig:ghoung}
  \end{figure}

Figure~\ref{fig:ghoung} serves as a visual representation, wherein the parameter space depicts the \textsf{QJD} of two stars, $S_1$ and $S_2$, as two concentric circles. Each circle's radius is proportional to the respective star's \textsf{QJD} value. The intersection of these circles at points $A$ and $B$ suggests that these coordinates could correspond to the North Celestial Pole for the epoch in question. When computing the parameters for any given pair of stars, the probability density for the actual North Celestial Pole is likely to be elevated at these intersection points. This intersection phenomenon underpins the core rationale of the Generalized Hough Transform algorithm.

However, the accurate determination of the true position is challenging due to observational errors, identification inaccuracies, and other variables. The use of Monte Carlo simulation provides a robust method for calculating these values. Utilizing the methodology of the Generalized Hough Transform, it is possible to determine the directional orientation of ancient instruments. However, this determined orientation does not entirely correspond to the actual direction of the North Celestial Pole. This discrepancy arises because the polar axes of ancient instruments typically fail to align precisely with the North Celestial Pole, resulting in observational data that often contain systematic errors. Nonetheless, by conducting a probabilistic analysis of the instrument's directional orientation, it is feasible to further infer the most probable location of the North Celestial Pole. Based on this inferred result, the specific observational epoch corresponding to the collected data can ultimately be ascertained.

The simulation method involves Gaussian sampling with a specified standard deviation (\textit{the standard deviation is measured in Chinese Degrees, where} $1~\textit{Chinese Degree} = \frac{360}{365.25} ~\textit{Degrees}$) for the \textsf{QJD}  data of each star. In this study, random sampling was performed 10,000 times. A sufficiently large number of simulations allows Monte Carlo simulation to produce precise results and provide confidence intervals or measures of uncertainty. Therefore, Monte Carlo simulation calculations can effectively reduce the impact of errors.

For each pair of stars, 10,000 simulations are conducted. After processing all data, the results are represented in the parameter space as a two-dimensional Gaussian distribution. By calculating the centroid of this distribution, we can ascertain the position of the North Celestial Pole during the observations.

It is crucial to consider the proper motion of stars based on potential epochs during the calculation process. All results are consistent with the J2000.0 coordinate system framework.

The Earth's precession, characterized by a cycle of approximately 26,000 years in which its axis of rotation shifts, allows us to employ state-of-the-art precession models \citep{2011A&A...534A..22V} to compute the theoretical position of the North Celestial Pole for any specified year across millennia. Through comparison of this computed position with the documented theoretical positions of the North Celestial Pole in historical records, we can infer the precise observational epoch of the data.

In a previous study \citep{hebl2024013}, the Generalized Hough Transform was utilized to analyze the star catalogs of the Song and Yuan dynasties. This approach exhibited a high degree of computational precision, effectively pinpointing the orientation of observational instruments for the respective eras and partially evaluating the observational inaccuracies of these instruments. For example, the observational error in the Song dynasty star catalog was roughly 0.3 Chinese Degrees, while in the Yuan dynasty, it was reduced to approximately 0.1 Chinese Degrees.

\section{Data analysis}

The \textit{Shi's Star Catalog} documents 120 stars, with 118 entries possessing full positional data (refer to Table~\ref{tab:ssxj-original}). The structuring of this data is grounded in pages 82-85 of the second edition of \textit{History of Chinese Stellar Observations} (\kaiti{中国恒星观测史}) \citep{PanNai2009HCSO}, drawing 114 stars from the \textit{Kaiyuan Zhanjing}. To account for the 6 absent stars from the \textit{Kaiyuan Zhanjing}, 4 were retrieved from a fragmentary manuscript of the \textit{Tiandi Ruixiang Zhi} (\kaiti{天地瑞祥志}), thereby compiling a comprehensive dataset of 118 stars.

In the initial determination of the epoch through the application of the generalized Hough transform, we utilized the identification data provided by \textit{Sun Xiaochun} in the study \textit{Research on the Shi Shi Constellations during the Han Dynasty} \citep{sunxiaochun1994ssxg,sun1997chinese}, encompassing the identification outcomes for 116 stars. However, the star \texttt{Langwei} \songti{郎位} is devoid of both RXD and QJD data. Consequently, the final dataset comprised 115 stars.

\begin{CJK}{UTF8}{gbsn}
\begin{table}
    \centering
    \scriptsize
    \caption{Shi's Star Catalog Original Data}
    \label{tab:ssxj-original}
    \begin{tabular}{|c|l|r >{\ttfamily}l|>{\ttfamily}c||c|l|r >{\ttfamily}l|>{\ttfamily}c|}
        \hline
        \textbf{No.} & \textbf{Constellation} & \multicolumn{2}{c|}{\textbf{RXD}}  & \multicolumn{1}{c||}{\textbf{QJD}} & \textbf{No.} & \textbf{Constellation} & \multicolumn{2}{c|}{\textbf{RXD}} & \multicolumn{1}{c|}{\textbf{QJD}} \\
        \hline
        1 & Jiao 角 & Jiao 角 & 00.00 & 091.00 & 61 & Tianda Jiangjun 天大将军 & Kui 奎 & 15.50 & 060.33 \\
2 & Kang 亢 & Kang 亢 & 00.00 & 089.00 & 62 & Daling 大陵 & Lou 娄 & 06.25 & 044.25 \\
3 & Di 氐 & Di 氐 & 00.00 & 094.00 & 63 & Tianchuan 天船 & Lou 娄 & 09.00 & 043.50 \\
4 & Fang 房 & Fang 房 & 00.00 & 108.00 & 64 & Juanshe 卷舌 & Wei 胃 & 10.25 & 056.00 \\
5 & Xin 心 & Xin 心 & 00.00 & 108.50 & 65 & Wuche 五车 & Bi 毕 & 03.00 & 063.00 \\
6 & Wei 尾 & Wei 尾 & 00.00 & 120.00 & 66 & Tianguan 天关 & Zi 觜 & 00.00 & 073.50 \\
7 & Ji 箕 & Ji 箕 & 00.00 & 118.00 & 67 & Nanhe 南河 & Jing 井 & 17.25 & 080.00 \\
8 & Dou 南斗 & Dou 斗 & 00.00 & 116.00 & 68 & Wuzhuhou 五诸侯 & Jing 井 & 02.00 & 057.00 \\
9 & Niu 牛 & Niu 牛 & 00.00 & 110.00 & 69 & Jishui 积水 & Jing 井 & 13.00 & 055.00 \\
10 & Nü 女 & Nü 女 & 00.00 & 106.00 & 70 & Jixin 积薪 & Jing 井 & 21.50 & 061.50 \\
11 & Xu 虚 & Xu 虚 & 00.00 & 104.00 & 71 & Shuiwei 水位 & Jing 井 & 19.50 & 072.50 \\
12 & Wei 危 & Wei 危 & 00.00 & 099.00 & 72 & Xuanyuan 轩辕 & Zhang 张 & 00.75 & 071.00 \\
13 & Shi 室 & Shi 室 & 00.00 & 085.00 & 73 & Shaowei 少微 & Zhang 张 & 10.50 & 070.50 \\
14 & Bi 壁 & Bi 壁 & 00.00 & 086.00 & 74 & Taiweiyuan 太微垣 & Yi 翼 & 09.00 & 076.50 \\
15 & Kui 奎 & Kui 奎 & 00.00 & 077.00 & 75 & Huangdizuo 黄帝座 & Yi 翼 & 09.50 & 063.50 \\
16 & Lou 娄 & Lou 娄 & 00.00 & 080.00 & 76 & Sidizuo 四帝座 &  &  &  \\
17 & Wei 胃 & Wei 胃 & 00.00 & 072.00 & 77 & Ping 内屏 & Yi 翼 & 07.00 & 072.50 \\
18 & Mao 昴 & Mao 昴 & 00.00 & 074.00 & 78 & Langwei 郎位 &  &  &  \\
19 & Bi 毕 & Bi 毕 & 00.00 & 078.00 & 79 & Langjiang 郎将 & Zhen 轸 & 08.00 & 039.25 \\
20 & Zi 觜 & Zi 觜 & 00.00 & 084.00 & 80 & Changchen 常陈 & Yi 翼 & 05.00 & 045.00 \\
21 & Shen 参 & Shen 参 & 00.00 & 094.50 & 81 & Santai 三台 & Jing 井 & 30.75 & 030.25 \\
22 & Jing 井 & Jing 井 & 00.00 & 070.00 & 82 & Xiang 相 & Yi 翼 & 05.00 & 031.50 \\
23 & Gui 鬼 & Gui 鬼 & 00.00 & 068.00 & 83 & Taiyangshou 太阳守 & Zhang 张 & 13.25 & 035.50 \\
24 & Liu 柳 & Liu 柳 & 00.00 & 079.00 & 84 & Tianlao 天牢 & Zhang 张 & 01.25 & 026.50 \\
25 & Xing 星 & Xing 星 & 00.00 & 091.00 & 85 & Wenchang 文昌 & Jing 井 & 15.75 & 025.75 \\
26 & Zhang 张 & Zhang 张 & 00.00 & 097.00 & 86 & Beidou 北斗 & Zhang 张 & 00.00 & 018.25 \\
27 & Yi 翼 & Yi 翼 & 00.00 & 099.00 & 87 & Ziweiyuan 紫微垣 & Zhen 轸 & 10.00 & 009.50 \\
28 & Zhen 轸 & Zhen 轸 & 00.00 & 099.00 & 88 & Gouchen 钩陈 & Bi 壁 & 08.75 & 011.50 \\
29 & Sheti 摄提 & Jiao 角 & 08.25 & 059.50 & 89 & Tianyi 天乙 & Zhen 轸 & 10.00 & 010.50 \\
30 & Dajiao 大角 & Kang 亢 & 02.50 & 058.00 & 90 & Taiyi 太乙 & Zhen 轸 & 10.00 & 010.00 \\
31 & Genghe 梗河 & Kang 亢 & 08.00 & 048.00 & 91 & Kulou 库楼 & Zhen 轸 & 00.25 & 140.00 \\
32 & Zhaoyao 招摇 & Di 氐 & 02.50 & 040.75 & 92 & Nanmen 南门 & Zhen 轸 & 14.00 & 130.00 \\
33 & Xuange 玄戈 & Di 氐 & 01.00 & 032.50 & 93 & Ping 平星 & Zhen 轸 & 14.00 & 100.00 \\
34 & Tianqiang 天枪 & Di 氐 & 00.75 & 028.75 & 94 & Qiguan 骑官 & Kang 亢 & 04.75 & 115.50 \\
35 & Tianbang 天棓 & Ji 箕 & 08.50 & 042.00 & 95 & Jizu 积卒 & Di 氐 & 13.75 & 124.25 \\
36 & NüChuang 女牀 & Ji 箕 & 01.00 & 050.00 & 96 & Gui 龟 & Wei 尾 & 12.00 & 131.00 \\
37 & Qigong 七公 & Di 氐 & 04.50 & 039.25 & 97 & Fuyue 傅说 & Wei 尾 & 12.75 & 120.50 \\
38 & Guansuo 贯索 & Wei 尾 & 00.50 & 059.25 & 98 & Yu 鱼 & Wei 尾 & 14.00 & 122.00 \\
39 & Tianji 天纪 & Wei 尾 & 05.00 & 051.50 & 99 & Chu 杵 & Ji 箕 & 01.75 & 132.50 \\
40 & Zhinü 织女 & Dou 斗 & 11.00 & 052.00 & 100 & Bie 鳖 & Dou 斗 & 01.00 & 129.50 \\
41 & Tianshiyuan 天市垣 & Wei 尾 & 00.75 & 094.25 & 101 & Jiukan 九坎 & Dou 斗 & 14.50 & 136.00 \\
42 & Dizuo 帝座 & Wei 尾 & 15.50 & 071.25 & 102 & Baijiu 败臼 & Nü 女 & 10.00 & 131.25 \\
43 & Hou 候 & Ji 箕 & 02.50 & 073.75 & 103 & Yulin 羽林军 & Wei 危 & 04.75 & 120.75 \\
44 & Huanzhe 宦者 & Wei 尾 & 12.00 & 072.50 & 104 & Beiluoshimen 北落师门 & Wei 危 & 09.00 & 130.75 \\
45 & Dou 斗 & Wei 尾 & 10.25 & 072.00 & 105 & Tusikong 土司空 & Bi 壁 & 07.75 & 120.25 \\
46 & Zongzheng 宗正 & Ji 箕 & 02.00 & 084.00 & 106 & Tiancang 天仓 & Kui 奎 & 04.75 & 120.00 \\
47 & Zongren 宗人 & Ji 箕 & 07.50 & 085.00 & 107 & Tianqun 天囷 & Wei 胃 & 06.25 & 096.50 \\
48 & Zong 宗 & Ji 箕 & 09.00 & 079.00 & 108 & Tianlin 天廪 & Wei 胃 & 11.25 & 090.00 \\
49 & Dongxian 东咸 & Xin 心 & 02.00 & 103.00 & 109 & Tianyuan 天苑 & Bi 毕 & 02.75 & 114.00 \\
50 & Tianjiang 天江 & Wei 尾 & 06.25 & 111.00 & 110 & Shenqi 参旗 & Bi 毕 & 09.50 & 093.00 \\
51 & Jian 建 & Dou 斗 & 07.25 & 113.25 & 111 & Yujing 玉井 & Bi 毕 & 12.25 & 102.75 \\
52 & Tianbian 天弁 & Dou 斗 & 06.75 & 090.75 & 112 & Ping 屏 & Zi 觜 & 00.75 & 118.00 \\
53 & Hegu 河鼓 & Dou 斗 & 22.75 & 085.00 & 113 & Ce 天厕 & Shen 参 & 03.25 & 115.00 \\
54 & Lizhu 离珠 & Nü 女 & 00.00 & 094.00 & 114 & Shi 屎 & Shen 参 & 07.00 & 123.00 \\
55 & Paogua 瓠瓜 & Nü 女 & 00.25 & 081.50 & 115 & Junshi 军市 & Jing 井 & 03.25 & 110.00 \\
56 & Tianjin 天津 & Dou 斗 & 02.00 & 049.00 & 116 & Yeji 野鸡 & Jing 井 & 08.00 & 111.00 \\
57 & Teshe 腾蛇 & Shi 室 & 01.50 & 051.00 & 117 & Lang 狼 & Jing 井 & 13.00 & 106.75 \\
58 & Wangliang 王良 & Bi 壁 & 00.50 & 042.50 & 118 & Hushi 弧矢 & Jing 井 & 16.00 & 122.75 \\
59 & Gedao 阁道 & Kui 奎 & 05.00 & 043.25 & 119 & Laoren 老人 & Jing 井 & 19.00 & 143.50 \\
60 & Fulu 附路 & Kui 奎 & 03.00 & 043.00 & 120 & Ji 稷 & Liu 柳 & 14.25 & 138.00 \\
        \hline
    \end{tabular}
\end{table}
\end{CJK}

Drawing upon the research outcomes of seminal scholars (\ref{sec:intro}), The year 1 BCE was initially designated as the reference epoch for the computational analysis. The dataset was subjected to Monte Carlo simulation to facilitate the generalized Hough transform analysis, thereby allowing for a preliminary estimation of the epochs. Subsequent analyses were then performed, grounded in these preliminary results.

\begin{figure}[htbp]
    \centering
    \includegraphics[width=0.5\textwidth]{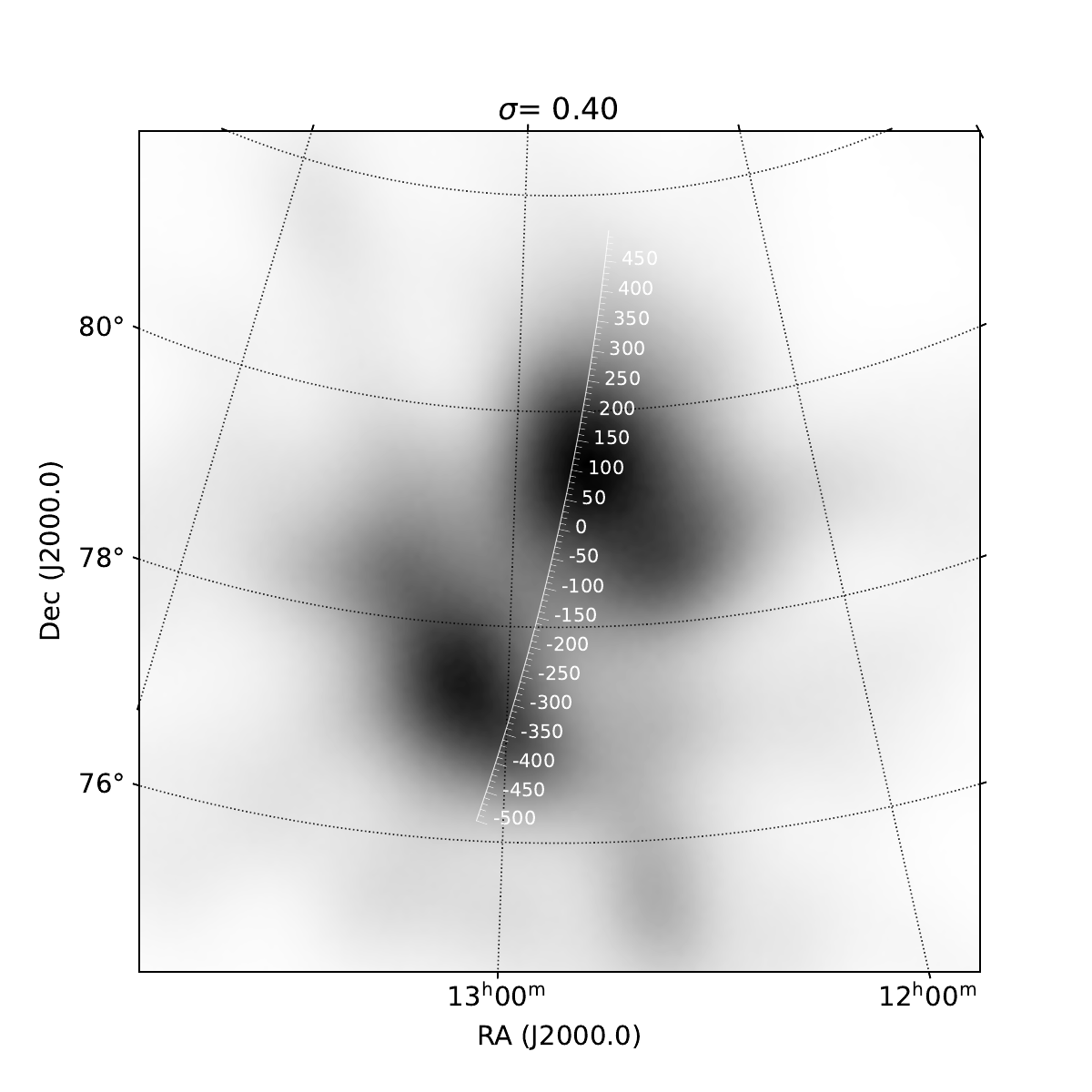}
    \caption{The calculation results using 1 BCE as the reference epoch.}
    \label{fig:ssxj-1}
  \end{figure}

The computational results, as depicted in Figure \ref{fig:ssxj-1}, reveal two distinct epochs with notable outcomes. A prominent peak is observed around 130 CE, accompanied by a smaller peak circa 350 BCE. These peaks are significantly separated, both temporally by approximately 500 years and spatially by about $2.8^\circ$. This indicates that the observational periods for the data within the \textit{Shi's Star Catalog} are clearly non-synchronous. The bulk of the observations are clustered around 130 CE, with an additional subset originating from approximately 350 BCE.

Within the dataset of these 118 stars (refer to Table \ref{tab:ssxj-original}), certain records exhibited substantial inaccuracies attributable to transcription errors in historical documentation. For instance, in Classical Chinese, the characters representing thirty (\songti{卅}) and forty (\songti{卌}) are visually similar, leading to potential errors during the copying of historical texts. Through meticulous calculations, analyses, and graphical comparisons, we have rectified these inaccuracies based on historical evidence. The rectified data encompasses:

\begin{itemize}
    \item \texttt{No. 31}, the \textsf{QJD} of \texttt{Genghe} (\songti{梗河}) should be 48 instead of 38. 
    \item \texttt{No. 35}, the \textsf{QJD} of \texttt{Tianbang} (\songti{天棓}) should be 42 instead of 32.
    \item \texttt{No. 55}, the \textsf{QJD} of \texttt{Paogua} (\songti{瓠瓜}) should be 81.5 instead of 71.5.
    \item \texttt{No. 111}, the \textsf{QJD} of \texttt{Yujing} (\songti{玉井})  should be 102.75 instead of 120.75. 
    \item \texttt{No. 119}, the \textsf{QJD} of \texttt{Laoren} (\songti{老人})  should be 143.5 instead of 133.5.
    \item \texttt{No. 120}, the \textsf{QJD} of \texttt{Ji} (\songti{稷}) should be 138 instead of 148.
\end{itemize}

Additionally, we rectified the identification data that exhibited significant deviations. Beyond the initial 115 stars, we successfully identified three previously unclassified stars (\texttt{Yu}~\songti{鱼}, \texttt{Gui}~\songti{龟}, and \texttt{Chu}~\songti{杵}) utilizing \textsf{RXD} and \textsf{QJD} data. Consequently, we compiled a comprehensive dataset of 118 stars with validated observational data.

Subsequent to these adjustments, calculations were conducted to assess the discrepancy between the theoretical and observed precession values at the two identified temporal points. This led to the selection of two distinct datasets, each correlated with one of the time periods: the first dataset included 59 stars with observations closer to the North Celestial Pole around 350 BCE, and the second dataset consisted of 59 stars with observations nearer to the North Celestial Pole around 130 CE. Each star was recalibrated according to its proper motion corresponding to the respective epoch.

\subsection{The First Epoch}

\begin{figure}[htbp]
    \centering
    \includegraphics[width=0.5\textwidth]{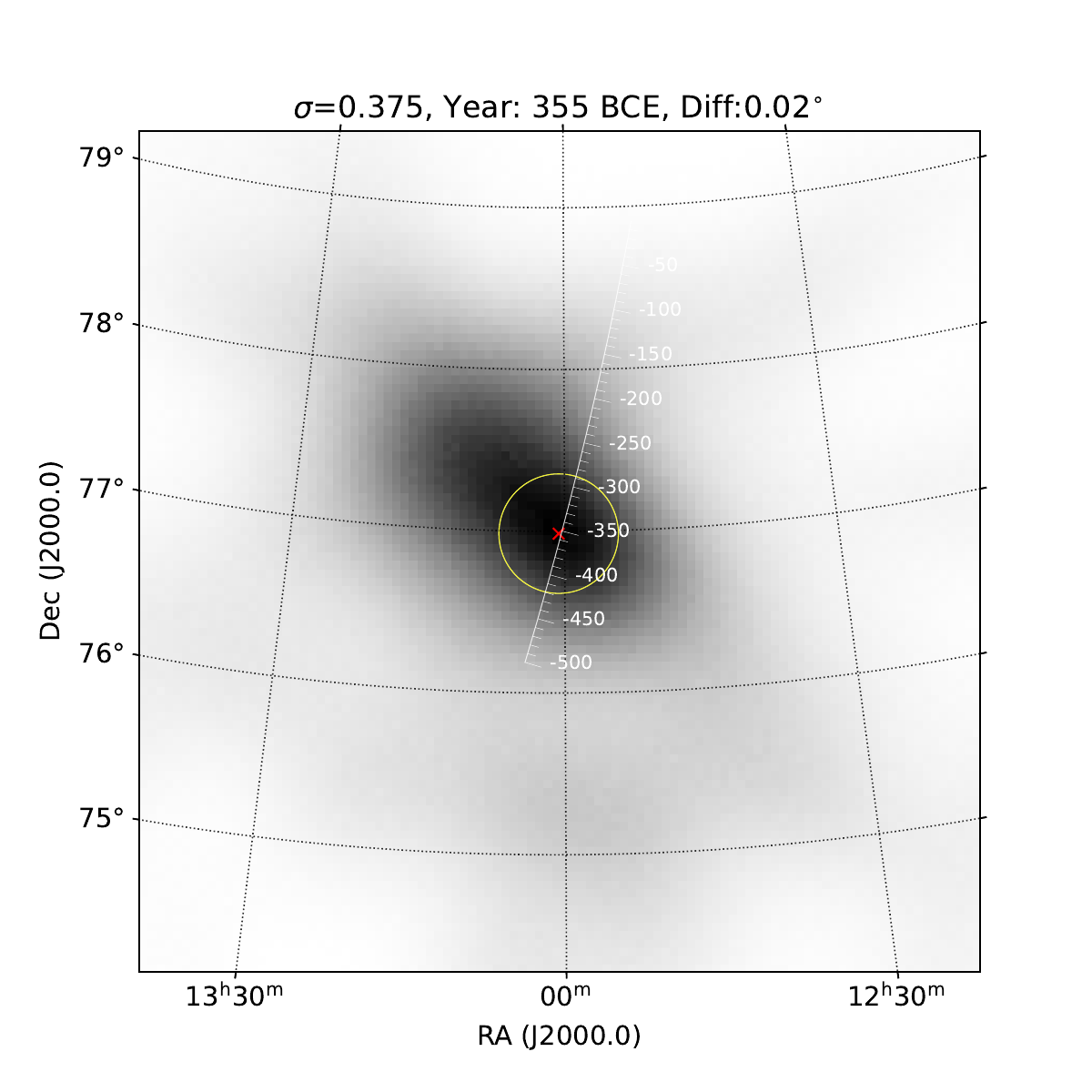}
    \caption{The epoch corresponding to the orientation of the first North Celestial Pole. (355 BCE)}
    \label{fig:ssxj-354}
  \end{figure}

An analysis of the 59-star candidate dataset from before 350 BCE, as depicted in Figure \ref{fig:ssxj-354}, involved Gaussian random sampling of each star 10,000 times with a standard deviation of $\sigma=0.375$ \textit{Chinese Degrees}, yielding an estimated epoch of approximately 355 BCE. The calculated centroid is roughly $0.02$ \textit{Chinese Degrees} from the North Celestial Pole's position in 355 BCE. The comprehensive dataset is presented in Table \ref{tab:ssxj-355}.

\begin{figure}[htbp]
    \centering
    \includegraphics[width=0.8\textwidth]{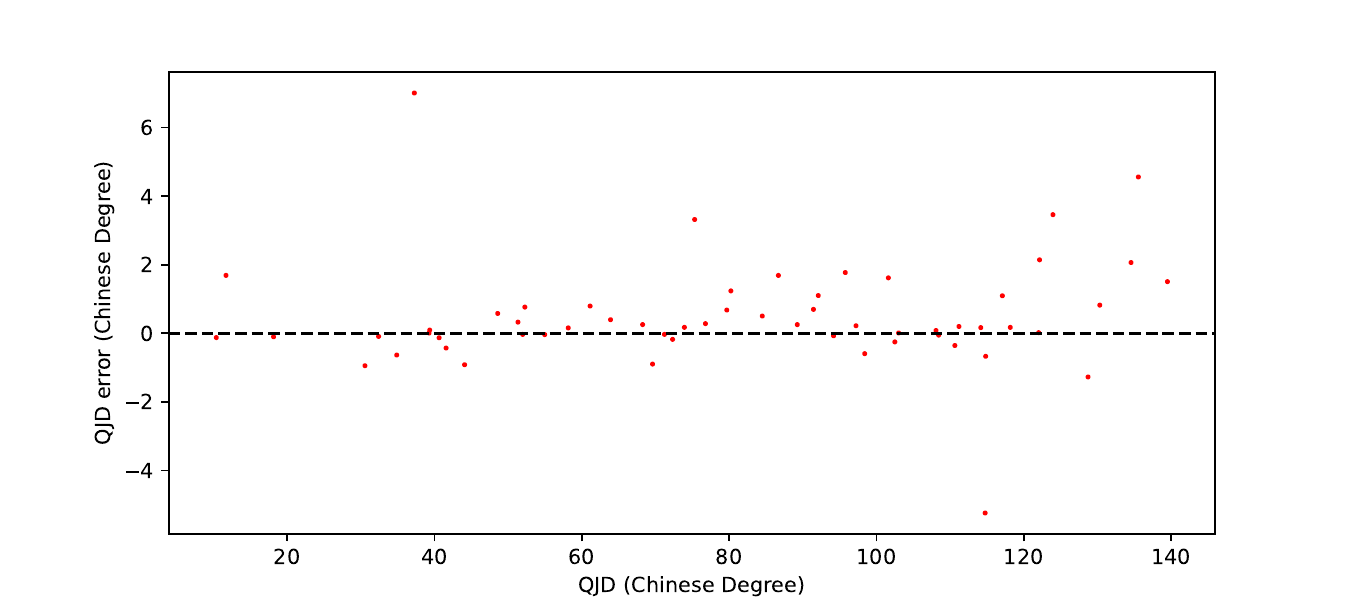}
    \caption{QJD errors in 355 BCE}
    \label{fig:ssxj-355-error}
\end{figure}

From the 59-star dataset, after excluding 6 stars with substantial errors, the remaining 53 stars exhibit an average discrepancy of $0.59$ \textit{Chinese Degrees} between the observed and theoretical declinations. The observed minimum deviation is $0.01$ \textit{Chinese Degrees}, and the maximum discrepancy reaches $2.14$ \textit{Chinese Degrees}. The results of the error analysis are provided in Figure \ref{fig:ssxj-355-error}.

\begin{CJK}{UTF8}{gbsn}
\begin{table}[htbp]
    \centering
    \scriptsize
    \caption{Shi's Star Catalog (355BCE)}
    \label{tab:ssxj-355}
    \begin{tabular}{|cl >{\ttfamily}c c >{\ttfamily}c >{\ttfamily}c >{\ttfamily}c >{\ttfamily}c >{\ttfamily}c|}
        \hline
        \textbf{No.} & \textbf{Xingguan} & \multicolumn{1}{c}{\textbf{HIP}} & \textbf{Star} & \multicolumn{1}{c}{\textbf{RA (355BCE)}}  & \multicolumn{1}{c}{\textbf{Dec (355BCE)}} & \multicolumn{1}{c}{\textbf{QJD}} & \multicolumn{1}{c}{\textbf{QJD\_err}} & \multicolumn{1}{c|}{\textbf{Pos\_err}} \\
        \hline
        2 & Kang 亢  & 69427 & $\kappa$ Vir & 12 11 20.37 & +02 01 45.58 & 89.00 & 0.25 & 0.25 \\
3 & Di 氐  & 72622 & $\alpha$ Lib & 12 46 30.54 & -04 23 45.09 & 94.00 & 1.77 & 1.75 \\
4 & Fang 房  & 78265 & $\pi$ Sco & 13 44 37.80 & -16 31 41.62 & 108.00 & 0.08 & 0.08 \\
5 & Xin 心  & 80112 & $\sigma$ Sco & 14 05 40.31 & -16 53 19.26 & 108.50 & -0.05 & 0.05 \\
6 & Wei 尾  & 82514 & $\mu$ Sco & 14 22 38.94 & -30 23 16.10 & 120.00 & 2.14 & 2.11 \\
7 & Ji 箕  & 88635 & $\gamma$ Sgr & 15 38 33.48 & -26 28 28.58 & 118.00 & 0.17 & 0.17 \\
8 & Dou 南斗  & 92041 & $\phi$ Sgr & 16 19 34.17 & -25 24 42.69 & 116.00 & 1.09 & 1.08 \\
23 & Gui 鬼  & 41822 & $\theta$ Cnc & 06 13 08.61 & +22 43 33.46 & 68.00 & 0.26 & 0.25 \\
24 & Liu 柳  & 42313 & $\delta$ Hya & 06 30 27.72 & +10 54 59.95 & 79.00 & 1.24 & 1.22 \\
25 & Xing 星  & 46390 & $\alpha$ Hya & 07 30 42.63 & -00 46 38.73 & 91.00 & 1.10 & 1.09 \\
26 & Zhang 张  & 48356 & $\upsilon$ Hya & 07 57 58.32 & -05 49 24.29 & 97.00 & 0.22 & 0.22 \\
27 & Yi 翼  & 53740 & $\alpha$ Crt & 09 06 25.45 & -06 59 31.94 & 99.00 & -0.59 & 0.58 \\
29 & Sheti 摄提  & 69727 & $\eta$ Boo & 12 15 58.39 & +05 42 19.20 & 59.50 & 26.02 & 26.03 \\
30 & Dajiao 大角  & 69673 & $\alpha$ Boo & 12 27 37.16 & +32 40 42.91 & 58.00 & 0.16 & 1.36 \\
31 & Genghe 梗河  & 71053 & $\rho$ Boo & 12 47 56.39 & +42 07 07.92 & 48.00 & 0.58 & 1.09 \\
32 & Zhaoyao 招摇  & 71075 & $\gamma$ Boo & 12 54 08.51 & +49 57 47.47 & 40.75 & -0.13 & 0.38 \\
33 & Xuange 玄戈  & 69732 & $\lambda$ Boo & 12 42 14.06 & +58 03 24.51 & 32.50 & -0.09 & 1.09 \\
35 & Tianbang 天棓  & 86414 & $\iota$ Her & 16 34 53.30 & +49 01 36.14 & 42.00 & -0.43 & 3.78 \\
37 & Qigong 七公  & 73555 & $\beta$ Boo & 13 32 00.46 & +51 13 13.61 & 39.25 & 0.10 & 4.34 \\
39 & Tianji 天纪  & 80181 & $\xi$ CrB & 14 51 22.80 & +38 29 08.01 & 51.50 & 0.77 & 1.91 \\
40 & Zhinü 织女  & 91262 & $\alpha$ Lyr & 17 17 52.12 & +38 46 44.83 & 52.00 & -0.03 & 2.91 \\
41 & Tianshiyuan 天市垣  & 81377 & $\zeta$ Oph & 14 31 48.55 & -02 49 25.16 & 94.25 & -0.07 & 1.55 \\
42 & Dizuo 帝座  & 84345 & $\alpha$ Her & 15 28 57.30 & +19 48 17.09 & 71.25 & -0.03 & 1.22 \\
43 & Hou 候  & 86032 & $\alpha$ Oph & 15 47 23.94 & +17 08 16.46 & 73.75 & 0.17 & 0.30 \\
44 & Huanzhe 宦者  & 83613 & 60 Her & 15 17 59.88 & +18 42 57.62 & 72.50 & -0.18 & 1.91 \\
45 & Dou 斗  & 83000 & $\kappa$ Oph & 15 08 15.86 & +15 45 47.30 & 72.00 & 3.32 & 3.50 \\
46 & Zongzheng 宗正  & 87108 & $\gamma$ Oph & 15 51 43.32 & +06 42 37.52 & 84.00 & 0.50 & 1.40 \\
47 & Zongren 宗人  & 88290 & 68 Oph & 16 03 57.54 & +04 33 27.33 & 85.00 & 1.69 & 1.96 \\
48 & Zong 宗  & 88765 & 71 Oph & 16 16 01.62 & +11 28 01.12 & 79.00 & 0.68 & 0.83 \\
49 & Dongxian 东咸  & 80343 & $\psi$ Oph & 14 12 34.12 & -11 31 45.93 & 103.00 & 0.01 & 0.24 \\
50 & Tianjiang 天江  & 84970 & $\theta$ Oph & 15 02 48.07 & -19 03 14.90 & 111.00 & -0.36 & 3.68 \\
52 & Tianbian 天弁  & 92117 & 5 Aql & 16 45 33.79 & -00 08 01.16 & 90.75 & 0.70 & 0.71 \\
57 & Teshe 腾蛇  & 111169 & $\alpha$ Lac & 20 59 20.20 & +39 24 41.01 & 51.00 & 0.33 & 2.90 \\
61 & Tianda Jiangjun 天大将军  & 9640 & $\gamma$ And & 23 53 45.72 & +29 45 12.43 & 60.33 & 0.80 & 1.25 \\
69 & Jishui 积水  & 37265 & $o$ Gem & 05 02 02.88 & +35 49 38.95 & 55.00 & -0.04 & 1.68 \\
73 & Shaowei 少微  & 52911 & 53 Leo & 08 40 33.78 & +21 23 51.55 & 70.50 & -0.90 & 0.93 \\
74 & Taiweiyuan 太微垣  & 57757 & $\beta$ Vir & 09 46 03.44 & +14 19 16.72 & 76.50 & 0.28 & 1.04 \\
75 & Huangdizuo 黄帝座  & 57632 & $\beta$ Leo & 09 43 47.79 & +27 01 16.19 & 63.50 & 0.40 & 0.39 \\
79 & Langjiang 郎将  & 63125 & $\alpha$ CVn & 10 55 31.65 & +51 18 24.43 & 39.25 & 0.01 & 0.96 \\
80 & Changchen 常陈  & 56997 & 61 UMa & 09 24 02.54 & +46 32 58.83 & 45.00 & -0.92 & 0.97 \\
81 & Santai 三台  & 44127 & $\iota$ UMa & 06 01 23.28 & +53 16 34.39 & 30.25 & 7.01 & 6.92 \\
82 & Xiang 相  & 57399 & $\chi$ UMa & 09 19 39.87 & +59 53 02.25 & 31.50 & -0.94 & 1.24 \\
83 & Taiyangshou 太阳守  & 54539 & $\phi$ UMa & 08 36 35.96 & +55 38 04.88 & 35.50 & -0.63 & 2.03 \\
86 & Beidou 北斗  & 54061 & $\alpha$ UMa & 07 47 00.35 & +72 06 43.36 & 18.25 & -0.10 & 0.85 \\
89 & Tianyi 天乙  & 62423 & 7 Dra & 10 31 27.72 & +79 46 24.90 & 10.50 & -0.12 & 1.16 \\
90 & Taiyi 太乙  & 63076 & 8 Dra & 10 49 50.45 & +78 28 39.32 & 10.00 & 1.69 & 1.70 \\
92 & Nanmen 南门  & 64004 & $\xi^2$ Cen & 11 05 22.20 & -36 52 37.45 & 130.00 & -1.27 & 1.97 \\
93 & Ping 平星  & 64962 & $\gamma$ Hya & 11 17 09.37 & -10 09 25.22 & 100.00 & 1.62 & 1.89 \\
94 & Qiguan 骑官  & 72010 & c1 Cen & 12 31 28.20 & -23 10 44.55 & 115.50 & -0.67 & 0.73 \\
96 & Gui 龟  & 85792 & $\alpha$ Ara & 14 42 33.85 & -43 36 44.46 & 131.00 & 4.56 & 6.82 \\
97 & Fushuo 傅说  & 87261 & G Sco & 15 15 47.45 & -32 10 44.70 & 120.50 & 3.46 & 3.47 \\
98 & Yu 鱼  & 87569 & M7 & 15 21 42.77 & -30 15 55.16 & 122.00 & 0.02 & 0.84 \\
99 & Chu 杵  & 90422 & $\alpha$ Tel & 15 37 11.07 & -42 37 50.44 & 132.50 & 2.06 & 2.56 \\
100 & Bie 鳖  & 92226 & $\mu$ CrA & 16 04 41.66 & -38 26 55.04 & 129.50 & 0.82 & 3.79 \\
106 & Tiancang 天仓  & 5364 & $\eta$ Cet & 23 08 11.18 & -23 06 35.16 & 120.00 & -5.24 & 5.17 \\
109 & Tianyuan 天苑  & 18543 & $\gamma$ Eri & 02 09 41.90 & -22 31 32.32 & 114.00 & 0.17 & 4.21 \\
111 & Yujing 玉井  & 23875 & $\beta$ Eri & 03 14 16.14 & -11 01 32.64 & 102.75 & -0.25 & 2.20 \\
116 & Yeji 野鸡  & 30324 & $\beta$ CMa & 04 40 00.12 & -19 36 02.80 & 111.00 & 0.20 & 1.41 \\
120 & Ji 稷  & 42913 & $\delta$ Vel & 07 38 45.66 & -47 30 06.24 & 138.00 & 1.51 & 2.55 \\
        \hline
    \end{tabular}
\end{table}   
\end{CJK}

\subsection{The Second Epoch}

\begin{figure}[htbp]
    \centering
    \includegraphics[width=0.5\textwidth]{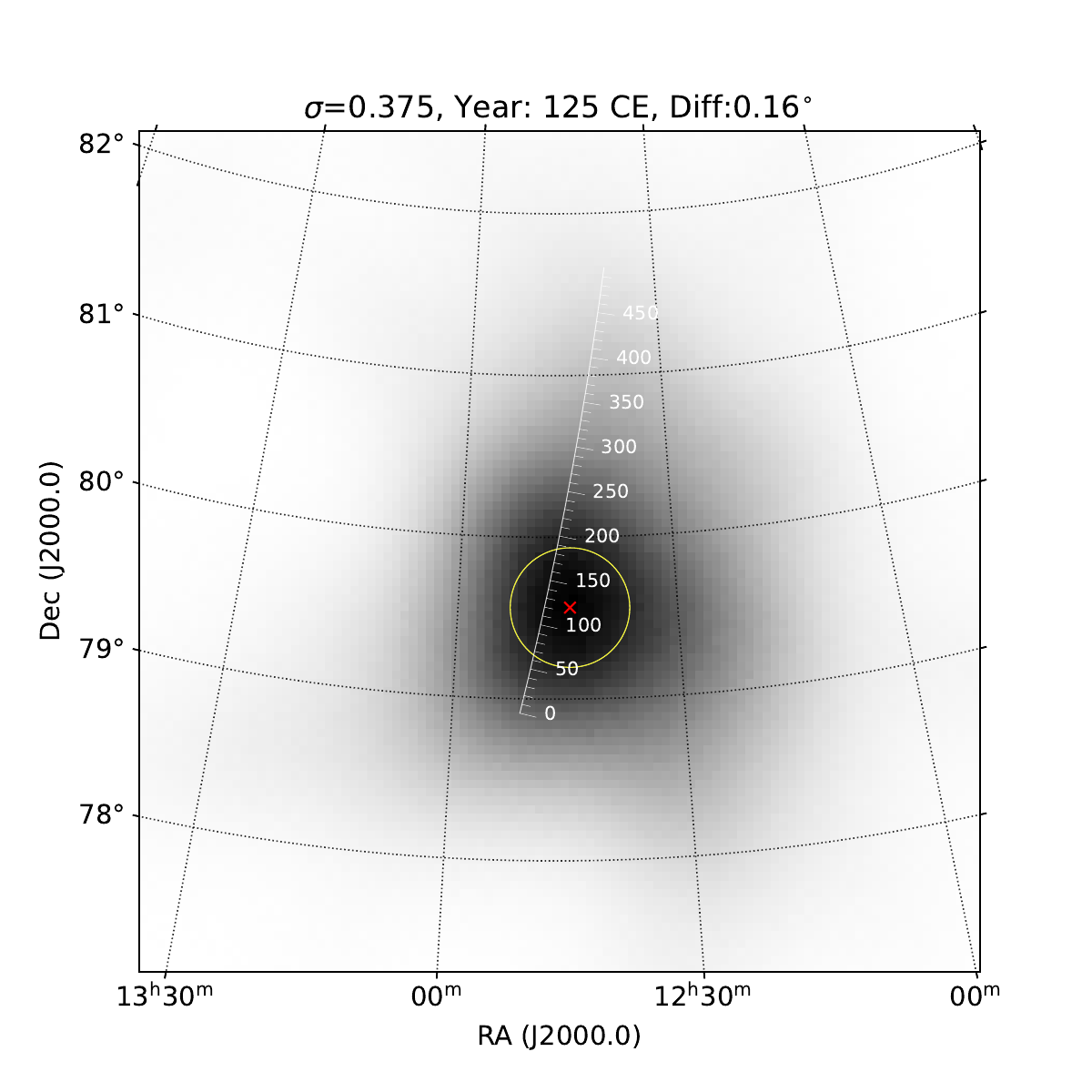}
    \caption{The epoch corresponding to the orientation of the second North Celestial Pole. (125 CE)}
    \label{fig:ssxj+125}
  \end{figure}

The analysis of the 59-star candidate dataset from the year 125 CE, as illustrated in Figure \ref{fig:ssxj+125}, involved Gaussian random sampling with a standard deviation of  $\sigma=0.375$ \textit{Chinese Degrees} for each star, conducted 10,000 times. This yielded a calculated observation epoch of approximately 125 CE. At this time, the computed centroid is roughly $0.16$ \textit{Chinese Degrees} distant from the North Celestial Pole's position in 125 CE. The full dataset is presented in Table \ref{tab:ssxj+125}.

\begin{figure}[htbp]
    \centering
    \includegraphics[width=0.8\textwidth]{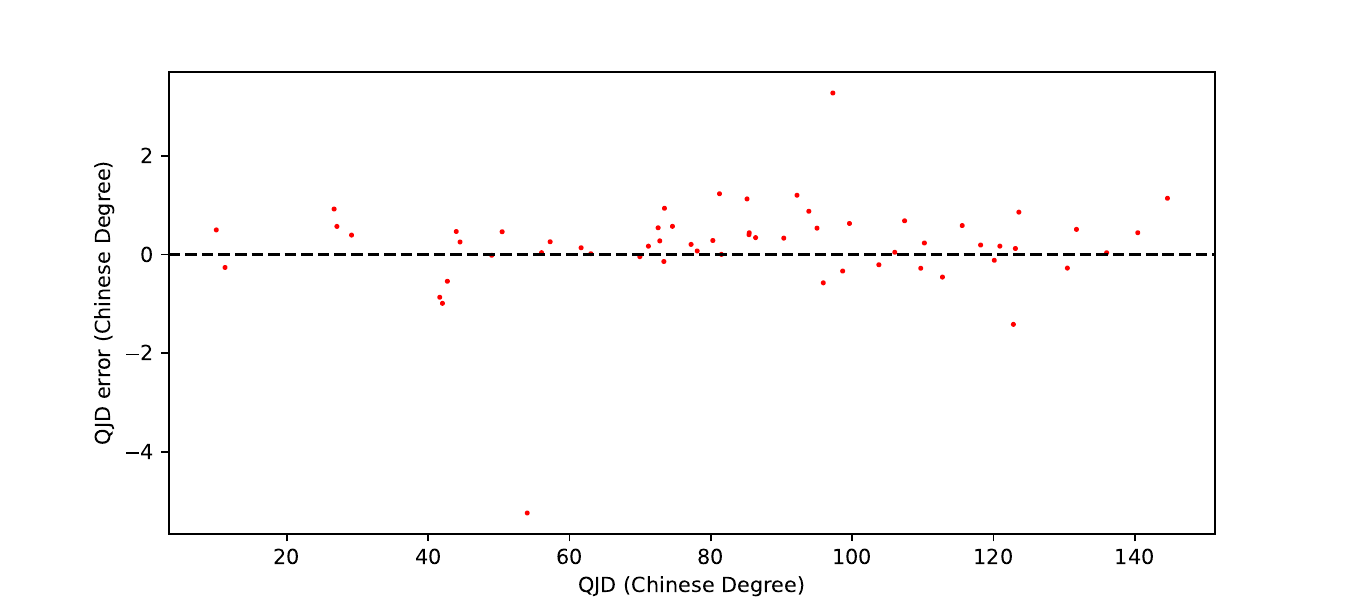}
    \caption{QJD errors in 125 CE}
    \label{fig:ssxj+125-error}
\end{figure}

Within the dataset of 59 stars, following the removal of 2 stars exhibiting significant deviations, the remaining 57 stars display an average discrepancy of $0.45$ \textit{Chinese Degrees} between the observed and theoretical declinations. The smallest recorded deviation is $0.001$ \textit{Chinese Degrees}, and the largest deviation extends to 1.42 \textit{Chinese Degrees}. The outcomes of the error analysis are presented in Figure \ref{fig:ssxj+125-error}.

\begin{CJK}{UTF8}{gbsn}

\begin{table}
    \centering
    \scriptsize
    \caption{Shi's Star Catalog (125CE)}
    \label{tab:ssxj+125}
    \begin{tabular}{|cl >{\ttfamily}c c >{\ttfamily}c >{\ttfamily}c >{\ttfamily}c >{\ttfamily}c >{\ttfamily}c|}
        \hline
        \textbf{No.} & \textbf{Constellation} & \multicolumn{1}{c}{\textbf{HIP}} & \textbf{Star} & \multicolumn{1}{c}{\textbf{RA (355BCE)}}  & \multicolumn{1}{c}{\textbf{Dec (355BCE)}} & \multicolumn{1}{c}{\textbf{QJD}} & \multicolumn{1}{c}{\textbf{QJD\_err}} & \multicolumn{1}{c|}{\textbf{Pos\_err}} \\
        \hline
        1 & Jiao 角  & 65474 & $\alpha$ Vir & 11 48 46.00 & -00 52 29.73 & 91.00 & 1.20 & 1.18 \\
9 & Niu 牛  & 100345 & $\beta$ Cap & 18 33 27.96 & -18 38 52.80 & 110.00 & 0.23 & 0.23 \\
10 & Nü 女  & 102618 & $\epsilon$ Aqr & 19 04 01.71 & -14 31 10.88 & 106.00 & 0.04 & 0.04 \\
11 & Xu 虚  & 106278 & $\beta$ Aqr & 19 50 42.00 & -12 17 56.34 & 104.00 & -0.21 & 0.21 \\
12 & Wei 危  & 109074 & $\alpha$ Aqr & 20 27 54.19 & -08 11 47.01 & 99.00 & 0.63 & 0.62 \\
13 & Shi 室  & 113963 & $\alpha$ Peg & 21 32 08.96 & +05 49 35.51 & 85.00 & 0.40 & 0.40 \\
14 & Bi 壁  & 1067 & $\gamma$ Peg & 22 38 31.23 & +04 53 59.41 & 86.00 & 0.34 & 0.34 \\
15 & Kui 奎  & 3693 & $\zeta$ And & 23 11 47.24 & +13 54 20.45 & 77.00 & 0.20 & 0.20 \\
16 & Lou 娄  & 8903 & $\beta$ Ari & 00 15 18.16 & +10 52 15.75 & 80.00 & 0.28 & 0.28 \\
17 & Wei 胃  & 12719 & 35 Ari & 00 59 18.86 & +18 30 07.33 & 72.00 & 0.54 & 0.53 \\
18 & Mao 昴  & 17499 & 17 Tau & 01 58 30.00 & +16 30 05.52 & 74.00 & 0.57 & 0.56 \\
19 & Bi 毕  & 20889 & $\epsilon$ Tau & 02 42 49.45 & +13 03 04.10 & 78.00 & 0.07 & 0.07 \\
20 & Zi 觜  & 26176 & $\phi$ Ori & 03 53 36.37 & +06 05 51.57 & 84.00 & 1.13 & 1.11 \\
21 & Can 参  & 25930 & $\delta$ Ori & 03 57 33.63 & -03 40 01.59 & 94.50 & 0.53 & 0.53 \\
22 & Jing 井  & 30343 & $\mu$ Gem & 04 30 29.67 & +21 03 04.10 & 70.00 & -0.05 & 0.04 \\
28 & Zhen 轸  & 59803 & $\gamma$ CrV & 10 41 35.32 & -07 14 43.51 & 99.00 & -0.34 & 0.33 \\
34 & Tianqiang 天枪  & 69483 & $\kappa$ Boo & 13 03 45.31 & +61 16 33.08 & 28.75 & 0.39 & 1.28 \\
36 & NüChuang 女牀  & 84380 & $\pi$ Her & 16 10 41.88 & +40 15 53.03 & 50.00 & 0.46 & 0.48 \\
38 & Guansuo 贯索  & 75695 & $\beta$ CrB & 14 10 38.52 & +36 46 08.66 & 59.25 & -5.24 & 10.17 \\
51 & Jian 建  & 93085 & $\xi$ Sgr & 17 05 42.35 & -21 10 12.03 & 113.25 & -0.46 & 2.75 \\
53 & Hegu 河鼓  & 97649 & $\alpha$ Aql & 18 18 56.35 & +05 47 15.63 & 85.00 & 0.44 & 0.45 \\
54 & Lizhu 离珠  & 101847 & 1 Aqr & 19 00 13.72 & -05 52 32.28 & 94.00 & 3.27 & 3.36 \\
55 & Paogua 瓠瓜  & 101769 & 6 Del & 19 09 24.38 & +09 40 13.20 & 81.50 & 0.00 & 1.08 \\
56 & Tianjin 天津  & 97165 & $\delta$ Cyg & 18 46 26.10 & +41 43 14.43 & 49.00 & -0.02 & 20.40 \\
58 & Wangliang 王良  & 746 & $\beta$ Cas & 22 40 21.81 & +48 58 00.64 & 42.50 & -0.87 & 0.86 \\
59 & Gedao 阁道  & 6242 & $\phi$ Cas & 23 37 22.02 & +47 54 27.18 & 43.25 & -0.54 & 1.12 \\
60 & Fulu 附路  & 4292 & $\upsilon$ Cas & 23 17 45.25 & +48 35 42.04 & 43.00 & -0.99 & 1.38 \\
62 & Daling 大陵  & 11060 & 9 Per & 00 27 52.43 & +46 08 13.19 & 44.25 & 0.25 & 2.10 \\
63 & Tianchuan 天船  & 13268 & $\eta$ Per & 00 50 40.13 & +46 40 01.16 & 43.50 & 0.46 & 0.46 \\
64 & Juanshe 卷舌  & 17529 & $\nu$ Per & 01 47 16.39 & +34 46 10.27 & 56.00 & 0.04 & 1.55 \\
65 & Wuche 五车  & 23015 & $\iota$ Aur & 02 59 56.48 & +27 53 28.13 & 63.00 & 0.01 & 1.17 \\
66 & Tianguan 天关  & 26451 & $\zeta$ Tau & 03 47 56.42 & +17 41 50.75 & 73.50 & -0.14 & 1.36 \\
67 & Nanhe 南河  & 36188 & $\beta$ CMi & 05 44 49.67 & +09 56 11.71 & 80.00 & 1.23 & 1.97 \\
68 & Wuzhuhou 五诸侯  & 33018 & $\theta$ Gem & 04 49 04.52 & +33 33 56.63 & 57.00 & 0.26 & 2.24 \\
70 & Jixin 积薪  & 38538 & $\phi$ Gem & 05 56 10.67 & +29 15 01.26 & 61.50 & 0.14 & 0.24 \\
71 & Shuiwei 水位  & 36760 & 68 Gem & 05 45 29.30 & +17 37 08.78 & 72.50 & 0.94 & 1.03 \\
72 & Xuanyuan 轩辕  & 49669 & $\alpha$ Leo & 08 25 19.38 & +19 51 18.92 & 71.00 & 0.17 & 0.31 \\
77 & Ping 内屏  & 57328 & $\xi$ Vir & 10 06 37.30 & +18 16 15.12 & 72.50 & 0.28 & 2.28 \\
84 & Tianlao 天牢  & 53261 & 44 UMa & 08 39 51.89 & +63 19 13.92 & 26.50 & 0.57 & 1.62 \\
85 & Wenchang 文昌  & 41704 & $o$ UMa & 05 38 29.66 & +63 42 44.95 & 25.75 & 0.92 & 1.11 \\
87 & Ziweiyuan 紫微垣  & 61281 & k Dra & 10 43 28.30 & +80 08 47.15 & 9.50 & 0.50 & 1.64 \\
88 & Gouchen 钩陈  & 11767 & $\alpha$ UMi & 23 18 27.50 & +78 55 32.26 & 11.50 & -0.26 & 0.37 \\
91 & Kulou 库楼  & 59747 & $\delta$ Cru & 10 46 39.54 & -48 25 18.98 & 140.00 & 0.44 & 0.81 \\
95 & Jizu 积卒  & 78384 & $\eta$ Lup & 14 03 44.11 & -31 04 00.74 & 124.25 & -1.42 & 1.43 \\
101 & Jiukan 九坎  & 98032 & $\iota$ Sgr & 17 41 17.89 & -44 04 45.06 & 136.00 & 0.03 & 0.83 \\
102 & Baijiu 败臼  & 107380 & $\iota$ PsA & 19 45 54.99 & -39 51 51.59 & 131.25 & 0.51 & 0.69 \\
103 & Yulin 羽林军  & 111449 & $\upsilon$ Aqr & 20 47 18.96 & -29 10 51.17 & 120.75 & 0.17 & 0.22 \\
104 & Beiluoshimen 北落师门  & 113368 & $\alpha$ PsA & 21 07 08.12 & -38 35 53.63 & 130.75 & -0.28 & 0.78 \\
105 & Tusikong 土司空  & 3419 & $\beta$ Cet & 23 06 51.79 & -28 24 16.79 & 120.25 & -0.12 & 0.50 \\
107 & Tianqun 天囷  & 14135 & $\alpha$ Cet & 01 26 34.52 & -04 32 44.35 & 96.50 & -0.58 & 0.86 \\
108 & Tianlin 天廪  & 15900 & o Tau & 01 46 50.28 & +00 58 03.91 & 90.00 & 0.33 & 0.86 \\
110 & Shenqi 参旗  & 22797 & $\pi^5$ Ori & 03 18 27.94 & -02 31 34.67 & 93.00 & 0.88 & 0.97 \\
112 & Ping 屏  & 23685 & $\epsilon$ Lep & 03 46 58.61 & -26 29 32.34 & 118.00 & 0.19 & 2.15 \\
113 & Ce 天厕  & 25606 & $\beta$ Lep & 04 08 44.95 & -23 55 28.17 & 115.00 & 0.59 & 0.69 \\
114 & Shi 屎  & 26460 & $\nu^2$ Col & 04 25 22.32 & -31 21 03.66 & 123.00 & 0.12 & 0.13 \\
115 & Junshi 军市  & 28816 & 17 Lep & 04 42 01.80 & -18 08 37.21 & 110.00 & -0.28 & 0.41 \\
117 & Lang 狼  & 32349 & $\alpha$ CMa & 05 22 28.13 & -15 53 16.98 & 106.75 & 0.68 & 0.69 \\
118 & Hushi 弧矢  & 32759 & $\kappa$ CMa & 05 40 15.53 & -31 49 51.77 & 122.75 & 0.86 & 1.66 \\
119 & Laoren 老人  & 30438 & $\alpha$ Car & 05 42 50.70 & -52 33 37.52 & 143.50 & 1.14 & 1.19 \\
        \hline
    \end{tabular}
\end{table}     
\end{CJK}

\section{Discussion and Conclusion}

The Generalized Hough Transform represents an advancement over prior methodologies by integrating a probabilistic computation model with the Monte Carlo algorithm for sampling. During the computation process, the model exclusively considers the \textsf{QJD}, while the \textsf{RXD} is excluded. This strategy effectively reduces the influence of various errors on the calculation outcomes, thereby significantly improving the accuracy of epoch estimation for observational data.

The application of the Generalized Hough Transform to the \textit{Shi's Star Catalog} data reveals that the catalog encompasses observations from two separate epochs. It is inferred that the \textit{Shi's Star Catalog} was originally compiled around 355 BCE, establishing a constellation system of 120 constellations and recording astronomical observations. Later, around 125 CE, an additional observation session was conducted, updating the data for 59 stars, which were then incorporated into the \textit{Shi's Star Catalog}, ensuring its preservation to the present day. 

The illustrious Eastern Han astronomer Zhang Heng (\kaiti{张衡})  assumed the exalted office of Grand Astrologer (\textit{Taishi Ling}, \kaiti{太史令}), a role entails the supervision of astronomical matters, on two separate occasions spanning the intervals of 115–120 CE and 126–132 CE. During his incumbency, he conceptualized and fabricated an armillary sphere(\textit{Hunyi}, \kaiti{浑仪}), which he employed to undertake methodical astronomical observations. Scholarly inquiry indicates that segments of the \textit{Shi's Star Catalog} may have been either revised or originally derived from the observational data amassed by Zhang Heng. Nevertheless, in the process of textual transmission across subsequent historical epochs, some data have preserved their original records dating back to the Zhanguo period, whereas others presumably embody observations conducted during Zhang Heng's own era. This dual provenance of the data—integrating antecedent records with contemporary observations—plausibly accounts for the coexistence of two discrete observational time frames within the \textit{Shi's Star Catalog}.

Our research indicates that the earliest recorded observations in the \textit{Shi's Star Catalog} date to approximately 355 BCE. In contrast, Japanese scholar Yasukatsu Motohashi has analyzed several ancient Western star catalogs \citep{1984Cent...27..280M}. The Timocharis catalog is from approximately 300 BCE to 250 BCE, the Aristyllus catalog from around 260 BCE, the Hipparchus catalog is believed to have originated around 130 BCE, and the Ptolemy catalog is dated to around 130 CE. Among these, Ptolemy's catalog is considered the most reliable and its epoch is relatively precise, indicating observations around 130 CE.

In summary, the computational method of the Generalized Hough Transform effectively mitigates the error issues encountered in previous methodologies, thus reliably resolving the challenge of determining the observation epochs of ancient Chinese stars. Consequently, the issue of observation epochs within the \textit{Shi's Star Catalog} is successfully addressed. In comparison to the observation epochs of other ancient star catalogs worldwide, the \textit{Shi's Star Catalog} predates even the oldest Western star catalogs, affirming its status as the oldest star catalog in the world.

In future studies, we will build upon the findings of this study to progressively reconstruct the complete dataset of the \textit{Shi's Star Catalog}, implement data visualization, and further advance related academic research.

\textbf{Acknowledgments}: This research has been significantly enriched by the foundational endeavors of numerous pioneering scholars. We are particularly indebted to Mr. Pan Nai~\citep{PanNai2009HCSO} and Mr. Sun Xiaochun~\citep{sun1997chinese} for their exceptional contributions to the compilation and identification of the data.

\bibliographystyle{raa}
\bibliography{ssxj.bib}

\end{document}